\documentclass[aps,prb,preprint,groupedaddress,showpacs]{revtex4}

\newcommand{\rbf}[0]{{\bf r}}

\usepackage{graphicx}

\begin{document}

\title{Electron and hole gas in modulation doped GaAs/AlGaAs radial heterojunctions}

\author{Andrea Bertoni}
\email{andrea.bertoni@nano.cnr.it}
\author{Miquel Royo}
\affiliation{S3, Istituto Nanoscienze - CNR, Via Campi 213/a, 41125 Modena, Italy}

\author{Farah Mahawish}
\author{Guido Goldoni}
\email{guido.goldoni@unimore.it}
\affiliation{Department of Physics, University of Modena and Reggio Emilia, Modena, Italy}
\affiliation{S3, Istituto Nanoscienze - CNR, Via Campi 213/a, 41125 Modena, Italy}

\date{\today}

\begin{abstract}
We perform self-consistent Schr\"odinger-Poisson calculations with exchange and correlation corrections to determine the electron/hole gas in a \emph{radial} hetero-junction formed in a modulation doped GaAs/AlGaAs core-multi-shell nanowire (CSNW) which is \emph{n-}/\emph{p-}doped. Realistic composition and geometry are mapped on an symmetry compliant two-dimensional grid, and the inversion/accumulation layers are obtained assuming mid-gap Fermi energy pinning at the surface. We show that the electron and hole gases can be tuned to different localizations and symmetries inside the core as a function of the doping level. Contrary to planar hetero-junctions, conduction electrons do not form a uniform 2D electron gas (2DEG) localized at the GaAs/AlGaAs interface, but rather show a transition between i) an isotropic, cylindrical distribution deep in the GaAs core (low doping), and ii) a set of six tunnel-coupled quasi-1D channels at the edges of the interface (high doping). Holes, on the other hand, are much more localized at the GaAs/AlGaAs interface and form either i) six separated planar 2DEGs at the GaAs/AlGaAs interfaces (low doping), ii) a quasi uniform six-fold bent 2DEG (intermediate doping), or iii) six tunnel-coupled quasi-1D channels at the edges (high doping). We also simulate the electron/hole gas in a CSNW-based field effect transistor. The field generated by a back-gate may easily deform the electron or hole gas, breaking the six-fold symmetry. Single 2DEGs at one interface or multiple quasi-1D channels are shown to form as a function of voltage intensity, polarity, and carrier type.
\end{abstract}


\maketitle

\section{Introduction \label{Introduction}}

\begin{figure}
 \includegraphics[width=0.45\textwidth]{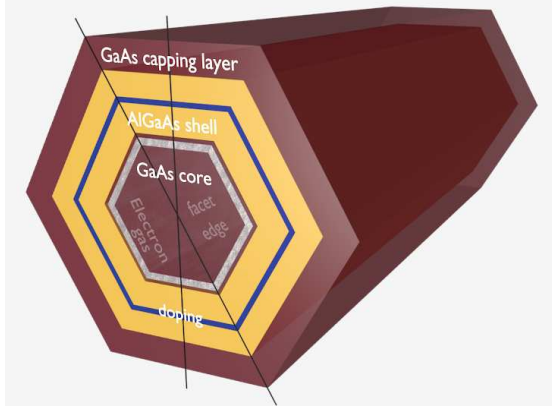}
 \caption{Schematics of the CSNW geometry. The free-carrier gas is formed in the GaAs core, eventually at the heterointerface with the AlGaAs shell. The shell is either  $n$- or $p$-doped in the middle with a 10~nm wide layer, and covered by a GaAs capping layer. The two thin lines crossing the hexagonal section represent the facet-to-facet and edge-to-edge directions on which the charge density is integrated in order to obtain the equivalent sheet charge density [see text and Eq.~(\ref{sheet_charge_density})]. Starting from the center and following the facet-to-facet direction, the core has a radius of 40~nm, while the thicknesses of the internal and capping layers are 50~nm and 10~nm, respectively.}
 \label{Geometry}
 \end{figure}

Semiconductor nanowires (NWs) are rapidly evolving into functional nano-materials\cite{Lieber03,Lieber07}. One key direction is the demonstration of core-multi-shell NWs (CSNWs), multi-layered materials where free-standing NWs, grown \emph{vertically} on top of a semiconductor surface, are used as a substrate for the \emph{radial} overgrowth of multilayers,\cite{Morral08,keplingerNL09,heigoldtJMC09,Sladek10}, strongly motivated by applications in energy harvesting\cite{Tian07,Colombo08,Czaban09} and electro-optical devices \cite{Tomioka10}. The resulting system is a radial hetero-structure, similar to 2D systems, like an hetero-junction (as in Fig.~\ref{Geometry}) or a quantum well, but grown radially, and wrapped around the core. Bridging between the self-assembling, bottom-up, quasi-1D nature of NWs and the flexible engineering of planar 2D hetero-structures, which opened up decades of astonishing developments in fundamental physics and innovative applications, CSNWs have the potential to bring the field of NW-based devices to the level of a new key nano-technology.

Critical steps have been taken in this direction. Long single-crystal, defect-free, untapered GaAs NWs, the material class of choice for quantum transport and high mobility, have been recently grown\cite{Shtrikman09a,Shtrikman09b}. Epitaxial GaAs/AlGaAs interfaces and multi-layered structures can be realized with high-quality interfaces\cite{Tambe08} and selective \emph{radial} doping has been demonstrated\cite{Tomioka10}. Self-catalytic or catalyst-free growth protocols have been developed, thus avoiding use of Au, a source of deep trap formation.\cite{heigoldtJMC09} High-mobility and ballistic transport, similarly to GaAs planar structures, are therefore realistic targets in CSNWs in the short run.

A typical geometry, as the one investigated in this paper, is shown in Fig.~\ref{Geometry}. GaAs NWs grow as hexagonal crystals along the [111] direction, exposing the six \{110\} facets. The GaAs core is overgrown by an epitaxial AlGaAs shell, possibly including a doping layer, and by a GaAs capping layer. Surface states easily deplete the outer layers, and a 2DEG may localize at the inner GaAs/AlGaAs hetero-interface, which consists of six planar \emph{facets} several tens of nm wide, interrupted by six \emph{edges}. The confined electronic system has the six-fold symmetry of the NW used as a substrate. 

It should be noted that in a typical hetero-junction the electron gas extends into the GaAs layer by tens of nm, due to the balance between kinetic and Coulomb energy scales. This is comparable to typical NWs diameters, and the electron gas on opposite sides of the GaAs core may strongly couple.\cite{Inoue85} Therefore, the free electron gas formation in a CSNW is a genuinely two-dimensional problem (we assume the CSNW to be translationally invariant in one dimension). For example, it is easy to infer that for large doping/size of the structure, Coulomb energy will induce charge to localize along six quasi-1D channels at the edges of the interface, to maximize the average electron-electron distance.

Ab-initio atomistic methods are computationally intensive and therefore limited to very thin NWs, ~1-2 nm in diameter. Nonetheless, they can provide important informations on the band-alignment and dopant properties.\cite{Zhang07,Amato11,Rosini11} Modelling of larger CSNWs is also limited. Recently, the single-particle properties of a \emph{cylindrical} electron gas have been simulated including  strong homogeneous and inhomogeneous magnetic fields\cite{Ferrari08B} and $k\cdot p$ effects.\cite{RaviKishore10}  Due to the prismatic geometry of CSNWs, however, cylindrical symmetry is broken, and localization at the edges of the hetero-interface may take place.\cite{FerrariNL,Masumoto10} Therefore, in NWs with diameters in the tens of nm range, the electronic system deviates substantially from the idealized cylindrical shape, and the electron gas may behave as a set of quasi-1D systems, rather than a uniform electron gas wrapped around the core. The above simulations neglect electron-electron interactions, though, and the confinement at the GaAs/AlGaAs interface is assumed, as, e.g., in a narrow radial quantum well. As discussed above, however, in doped structure the in-plane directions cannot be disentangled and full calculations need to be performed where the confinement is self-consistently included.\cite{wongNL11}

In this paper we discuss the formation of the electron and hole gas in GaAs/AlGaAs core-shell NWs which are \emph{n-} or \emph{p-}doped in the AlGaAs barriers. We will show that remote charge density which forms at the inner GaAs/AlGaAs interface may be tuned with doping and/or external gates changing the symmetry and localization of the distribution. The electron/hole gas has a mixed dimensionality, which is locally 1D or 2D. Remarkable difference, from this point of view, is found between electron and hole gases. We finally show that the electron or hole gas can be easily reshaped and tuned between 2D and 1D channels, which can be individually depleted, by a transverse electric field applied by a back-gate.

The following of this paper is organized as follows. In the next Sec.~\ref{Method}, the physical and numerical modelling is sketched, and the self-consistent procedure is described. In Sec.~\ref{Results}, the results of our calculations are illustrated. First, we introduce the prototypical structure that we investigate and the relevant simulative parameter. In Subsections~\ref{Electrons} and \ref{Holes}, we report our results for $n$-doped and $p$-doped systems, respectively. The origin of the localization patterns will be discussed in Subsection~\ref{Patterns}, while in Subsection~\ref{Bias} will deal with a CSNW subject to an external electric field in a field effect transistor configuration. Finally, in Sec.~\ref{Conclusions} we draw our conclusions.

\section{Method \label{Method}}

The free electron/hole gas of the CSNW is obtained within an envelope-function approach in a single-band approximation, including carrier-carrier interaction at a mean-field level by solving the Poisson equation for the free charge distribution plus the charge due to the remote doping. To solve the coupled system of Schr\"odinger and Poisson equations we adopt the usual self-consistent approach. Assuming translational invariance along the NW growth axis $z$, we need to solve numerically a 2D problem in the $x,y$ plane. At each iteration, we first solve the effective-mass Schr\"odinger equation on a 2D hexagonal domain, representing the cross-section of the CSNW, for both electron and holes,
\begin{equation} \label{schroed2Delec}
\left[-\frac{\hbar^2}{2}\nabla_\rbf
\left(\frac{1}{m_e(\rbf)} \nabla_\rbf \right)
+ E_c(\rbf) - eV(\rbf) \right] \phi^e_n(\rbf) =
\epsilon^e_n \phi^e_n(\rbf) \, ,
\end{equation}
\begin{equation} \label{schroed2Dhole}
\left[-\frac{\hbar^2}{2}\nabla_\rbf
\left(\frac{1}{m_h(\rbf)} \nabla_\rbf \right)
- E_v(\rbf) + eV(\rbf) \right] \phi^h_n(\rbf) =
-\epsilon^h_n \phi^h_n(\rbf) \, ,
\end{equation}
where $\rbf = (x,y)$ is the 2D coordinate, $m_e$ and $m_h$ are the material-dependent effective masses of electrons and holes, assumed isotropic in the $x,y$ plane, $E_c(\rbf)$ and $E_v(\rbf)$ are the local conduction-band and valence-band edges, respectively, and $V(\rbf)$ is the electrostatic potential generated by free carriers and fully-ionized dopants. The above equations are numerically integrated to give the set of eigenfunctions (eigenenergies) for electrons and holes $\phi^e(\rbf)$ ($\epsilon^e$) and $\phi^h(\rbf)$ ($\epsilon^h$), respectively.

After the solution of Eqs.~(\ref{schroed2Delec}) and (\ref{schroed2Dhole}), we calculate the \emph{volumetric} total charge density
\begin{equation} \label{totchargedensity}
\rho(\rbf) =  e\big(n_h({\rbf})-n_e({\rbf})+\rho_D(\rbf)-\rho_A(\rbf)\big) ,
\end{equation}
where $\rho_D$ and $\rho_A$ are the ionized donor and acceptor densities, respectively,
and the densities of free electrons and holes are
\begin{equation} \label{elecdensity}
n_e(\rbf) =  2 \sum_n |\phi^e_n(\rbf)|^2 \, \sqrt{\frac{\overline{m}_e(\rbf)k_BT}{2\pi\hbar^2}}
\mathcal{F}_{-\frac{1}{2}}\left(\frac{-\epsilon^e_n+\mu}{k_B T}\right) \, ,
\end{equation}
\begin{equation} \label{holedensity}
n_h(\rbf) =  2 \sum_n |\phi^h_n(\rbf)|^2 \, \sqrt{\frac{\overline{m}_h(\rbf)k_BT}{2\pi\hbar^2}}
\mathcal{F}_{-\frac{1}{2}}\left(\frac{\epsilon^h_n-\mu}{k_B T}\right) \, .
\end{equation}
Here, $\overline{m}_e$ ($\overline{m}_h$) is the effective electron (hole) mass along the CSNW axial direction (which is in general different from the in-plane mass in Eqs.~\ref{schroed2Delec},\ref{schroed2Dhole}, particularly for holes), $T$ is the temperature, $\mu$ is the Fermi level, and $\mathcal{F}_{k}(x)=\frac{1}{\Gamma(k+1)}\int_0^\infty\frac{t^k dt}{e^{t-x}+1}$ is the complete Fermi-Dirac integral of order $k$, which results from the integration of the parabolic dispersion along the growth direction $z$.

From the total charge density, we compute the
electrostatic potential $V$ solving the Poisson equation with a material-dependent relative dielectric constant $\varepsilon_r$
\begin{equation}\label{poissoneq}
\nabla_\rbf \left[ \varepsilon_r(\rbf) \nabla_\rbf \, V(\rbf) \right]
= - \frac{\rho(\rbf)}{\varepsilon_0} \, .
\end{equation}
We use Dirichlet boundary conditions, with the potential on the domain boundaries fixed to either an input value representing the voltage of a metallic gate in that position, or zero. Several test simulations, with increasing domain size, confirm that the electrostatic potential at the CSNW surface is essentially zero when no gate is included.

The contribution due to exchange and correlation effects is calculated via a local density approximation from the densities $n_e(\rbf)$ and $n_h(\rbf)$, and added to $V$.  We adopt the well known approximate expression reported in Ref.~\onlinecite{gunnarssonPRB76} and \onlinecite{andoJAP03}. In all the simulations presented in the following, it turns out to be almost negligible\cite{wongNL11}. For the sake of brevity, here we consider the exchange-correlation contribution already included in $V$.

The whole procedure is then iterated, with the new potential $V$ inserted in Eqs.~(\ref{schroed2Delec}) and (\ref{schroed2Dhole}).  The self-consistent cycle is stopped when the relative variation of the charge density is lower than a given value $E_\rho$ at any position, i.e.,
\begin{equation}\label{StoppingCrit}
    \frac{A \max|\rho(\rbf)-\rho'(\rbf)| }{\int_A\rho(\rbf) d\rbf} < E_\rho ,
\end{equation}
where $A$ is the area of the simulation domain.

Once the convergence is achieved, one can calculate the linear charge density (we recall that the system is assumed to be translationally invariant along $z$) by the area integration
\begin{equation}\label{eq:linear_charge_density}
    \rho_{\mbox{\scriptsize linear}} = \int_A \rho(\rbf) d\rbf.
\end{equation}
To quantify the charge distribution in the section of the CSNW structure, we also define the \emph{effective} sheet charge density
\begin{equation}\label{sheet_charge_density}
    \rho_{\mbox{\scriptsize sheet}} = \frac{1}{2} \int_L \rho [\mathbf{r}(l)] dl
\end{equation}
where $L$ is one of the two directions shown in Fig.~\ref{Geometry}, which either join two edges or two facets, and $\mathbf{r}(l)$ is a point on one of these lines.  $\rho_{\mbox{\scriptsize sheet}}$ represents the sheet charge of an equivalent \emph{planar} heterojunction, as the charge were uniform along the interface. In our case, $\rho_{\mbox{\scriptsize sheet}}$ is in general different in the two directions, and is an indication of the preferential localization of the free charge.

With regards to the numerical approach, the solutions of Schr\"odinger and Poisson equations are obtained through a box integration method\cite{selberherr84}, on a triangular grid with hexagonal elements. This choice reproduces the shape of the domain and the symmetry of the (unbiased) system, and avoids numerical artifacts originated by discretization asymmetries of the six domain boundaries, as would be the case, e.g, using a rectangular grid. Formally, the partial differential equations Eqs.~(\ref{schroed2Delec}), (\ref{schroed2Dhole}) and (\ref{poissoneq}) are integrated on each hexagonal element. By applying the divergence theorem, the area integral is converted in a linear integral of the flux along the hexagon boundary. A balance between incoming and outgoing fluxes (obtained through a first-order finite-differences scheme) of adjacent hexagons connects the unknowns on different elements. This results in a symmetric (Hermitian) sparse matrix for the Poisson (Schr\"odinger) equation, whose dimension correspond to the number of hexagons, and with six non-zero elements on each row. The matrix for the Schr\"odinger equation is diagonalized through a Lanczos library algorithm\cite{Arpack98} while that for the Poisson equation is solved, with the known term obtained from $\rho$, via an efficient library routine\cite{Pardiso08}. The Fermi-Dirac integral is computed by using an efficient routine implementing Chebyshev polynomial expansion\cite{MacLeod98}.

Stability of the self-consistent cycle for large and/or heavily doped structures is delicate. The usual mixing procedure using an under-relaxation parameter, $\alpha$, is implemented: at iteration $i$, the electrostatic potential at the current iteration $V^i$ is substituted by
\begin{equation}\label{eq:underelaxation}
    V^i \leftarrow \alpha V^i + (1-\alpha) V^{i-1}.
\end{equation}
In addition, although the Poisson and Schr\"odinger equations are solved on the same grid points, the charge density extends very little into the AlGaAs barriers, due to the large depletion operated by surface states. Therefore, the Schr\"odinger equation need to be solved in a much smaller region. Typically we require the wave functions to vanish at the position of the doping layer. 

Holes are treated in a single-band approach. As we shall discuss in Sec.~\ref{Holes}, the hole gas is mostly confined at the GaAs/AlGaAs interfaces, which are directed along the six \{110\} directions. The parabolic dispersion for heterostructures grown along the [110] direction is given in \cite{Fishman95}, Eq.~4.4a. The strongly anisotropic mass is heavy in the [110] direction (roughly twice the mass along the widely used [100] direction) and light along in-plane direction, here corresponding to the vertical growth direction [111], as shown later in Table \ref{table_mat_par}.

\section{Results \label{Results}}

\subsection{Structure and simulation details \label{Structure}}

As a prototype structure (see Fig.~\ref{Geometry}), we simulate a device similar to the one described in Ref.~\onlinecite{Sladek10}.  A GaAs core 80 nm wide is surrounded by a 50 nm wide Al$_{0.3}$Ga$_{0.7}$As shell, and by a 10 nm wide GaAs capping layer. The Al$_{0.3}$Ga$_{0.7}$As shell is uniformly doped in the center with a 10 nm wide layer at a constant doping density $\rho_D$. Material parameters are shown in Tab.~\ref{table_mat_par}. $\mu$ is taken at the mid-gap value of GaAs, and all calculations are performed at $T=20K$.

 \begin{table}
 \begin{tabular}{lcc}
  \hline
        & GaAs & Al$_{0.3}$Ga$_{0.7}$As \\
  $E_c-E_v$ [eV]     & 1.43 & 1.858 \\
  $\Delta E_c$ [eV]  &  \multicolumn{2}{c}{0.284}  \\
  $\Delta E_v$ [eV]  &  \multicolumn{2}{c}{0.144}   \\
  $m_e$               & 0.067 & 0.092 \\
  $\overline{m}_e$   & 0.067 & 0.092 \\
  $m_h$                  & 0.690 & 0.731 \\
  $\overline{m}_h$       & 0.105 & 0.124 \\
  $\varepsilon_r$        & 13.18 & 12.24 \\
  \hline
 \end{tabular}
 \caption{Material parameters used in the simulations. $\Delta E_c$ and $\Delta E_v$ are the conduction and valence band offsets, respectively, between the conduction band edge $E_c$ and valence band edge $E_v$ of the GaAs and the Al$_{0.3}$Ga$_{0.7}$As layers.\cite{Bosio88} $m_e$ and $\overline{m}_e$ are the relative electron effective masses in the $x,y$ plane, orthogonal to the CSNW axis, and along the axis direction, $z$, respectively. Analogously for $m_h$ and $\overline{m}_h$ for holes. Note that the electron effective mass is assumed isotropic, while hole relative effective masses are strongly anisotropic.}
 \label{table_mat_par}
 \end{table}

In this Section we report results for two kind of structures, one $n$-doped and one $p$-doped. In both cases, the occupation probability of minority carriers is negligible, hence, we need to solve only the relevant equation, among Eq.~(\ref{schroed2Delec}) and (\ref{schroed2Dhole}). Schr\"odinger and Poisson equations are solved on a 2D hexagonal domain, discretized with a regular hexagonal tessellation. Our simulations typically use about $2\times10^{5}$ hexagonal elements. The Schr\"odinger equation is solved on a sub-set of $5\times10^{4}$ elements, as explained in the previous section. The under-relaxation parameter $\alpha \sim 0.05$ and the maximum convergence relative error allowed in the charge density $E_\rho = 10^{-3}$.
Convergence is typically achieved within 100$\div$200 iterations if the potential $V$ at the first iteration is zeroed. Clearly, the number of iterations can be substantially reduced by initializing the potential with a suitable guess, typically, the converged potential at a slightly different doping density. Converged results, of course, do not depend on this choice.

\subsection{Electron gas localization \label{Electrons}}

\begin{figure}
 \includegraphics[width=0.9\textwidth]{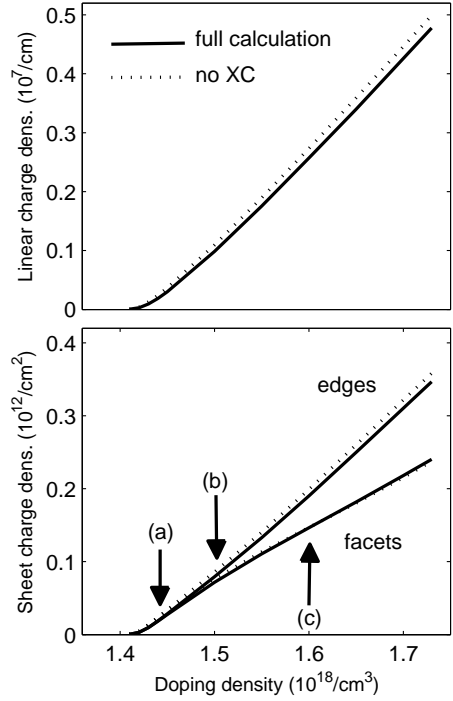} 
 \caption{Top: Linear charge density of the free-electron gas as a function of the donor density $\rho_D$. Bottom: sheet charge density of the free-electron gas calculated along the facet-to-facet and edge-to-edge directions indicated in Fig.~\ref{Geometry}.
 Arrows indicate three different densities shown in in Fig.~\ref{electron_densities}
 As the doping density increases, the latter dominates, indicating that the free electrons tend to accumulate on the six CSNW edges and form six 1D channels. In both graphs, the solid (dotted) curves represent the results of self-consistent calculations including (neglecting) exchange-correlation contributions.
}
 \label{electron_free_charge_density}
 \end{figure}

We now consider a $n$-doped CSNW, where a free-electron gas is eventually formed in the GaAs core or at the inner heterointerface, and specifically investigate the free electron gas formation as a function of the doping density $\rho_D$. Figure \ref{electron_free_charge_density} (top panel) shows the linear free electron density vs $\rho_D$. At a threshold density $\sim 1.4\times 10^{18} cm^{-3}$ the nanowire starts to be populated, in qualitative agreement with \onlinecite{Tomioka10,Sladek10}. The linear density is a fraction of $\sim 10^7 cm^{-1}$, and increases linearly with doping. Note the tiny difference between the calculations with and without the XC potential. In the rest of the paper, all results are reported for calculations which include the XC potential.

\begin{figure}
 \includegraphics[width=0.45\textwidth]{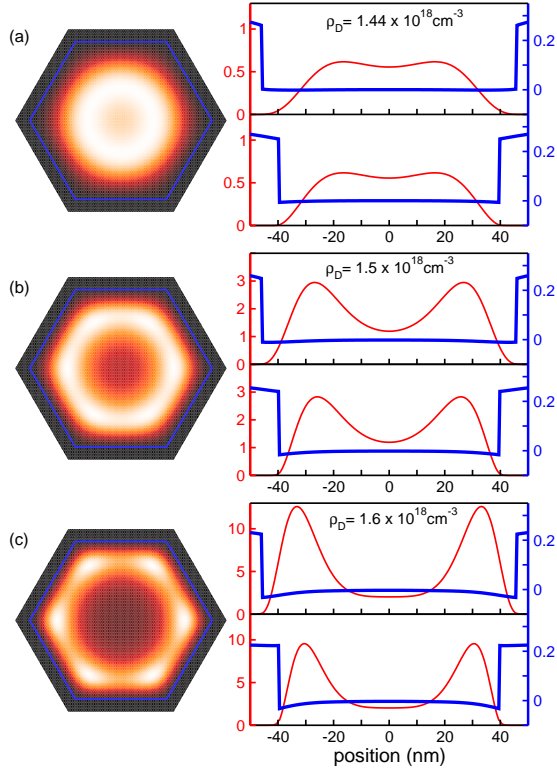}
 \caption{Spatial distribution of the free-electron gas for the three values of the doping density indicated as (a), (b), and (c) in Fig.~\ref{electron_free_charge_density} (bottom panel). For each density we show (left) the 2D map of the charge, with the GaAs/AlGaAs interface indicated as a solid line, and (right) the charge density profile (thin line, left axis, units of $10^{15}$~cm$^{-3}$) and the self-consistent conduction-band profile (thick line, right axis, units of eV, $\mu=0$) along the edge-to-edge direction (right, top) and along the facet-to-facet (right, bottom). Note the different scale for different doping density. }
 \label{electron_densities}
 \end{figure}

A glimpse to the charge density maps reported in Fig.~\ref{electron_densities} shows that the free electron gas is localized in the GaAs core, as expected. Indeed, at sufficiently large doping, the conduction band bends down from the surface value until the Fermi level (zero in the right energy scale of the right plots of Fig.~\ref{electron_densities}), which is pinned to the middle of the gap at the CSNW surface, is situated just above the GaAs conduction edge of the core. This is in complete analogy to the formation of an inversion layer in a remotely doped planar heterojunction. However, here the localization is not homogeneous along the heterointerface, and its pattern strongly depends on the doping density, as we discuss below.

In Fig.~\ref{electron_free_charge_density} (bottom panel) we show the sheet charge density along facets and edges. At small doping the two densities are equal, while at larger doping the edge population becomes dominant. This evolution is made more clear in the charge density maps of Fig.~\ref{electron_densities}, which refer to the three densities labelled (a), (b) and (c) in Fig.~\ref{electron_free_charge_density}.
At the lowest doping (Fig.~\ref{electron_densities}(a), when the GaAs core starts to be populated, the charge is distributed deep into the core. The distribution is only slightly modulated (right panels) crossing the core along either the facet-to-facet of edge-to-edge directions, and slightly depleted in the center of the core. Furthermore, the distribution is almost circularly symmetric. Accordingly, the two sheet charge densities in Fig.~\ref{electron_free_charge_density} are equal.

As the doping is increased [Fig.~\ref{electron_densities}(b)] the charge depletion in the center is more pronounced, and charge moves toward the interfaces. The 2D maps (left panel) shows that the distribution is starting to develop a six-fold symmetry, and the sheet charge density along the edge-to-edge direction is starting to dominate. For large doping [Fig.~\ref{electron_densities}(c)] the charge is strongly localized at the edges. This can be seen in the 2D maps (left panel) as well as, more quantitatively, in the two profiles (right panels). In this regime the charge density has fully developed the six-fold symmetry and resemble a set of coupled quantum wires more than a 2DEG, since most of the charge is confined in relatively narrow channels at the edges, while only a minor part of the charge sits at the facets.

\begin{figure}
 \includegraphics[width=0.9\textwidth]{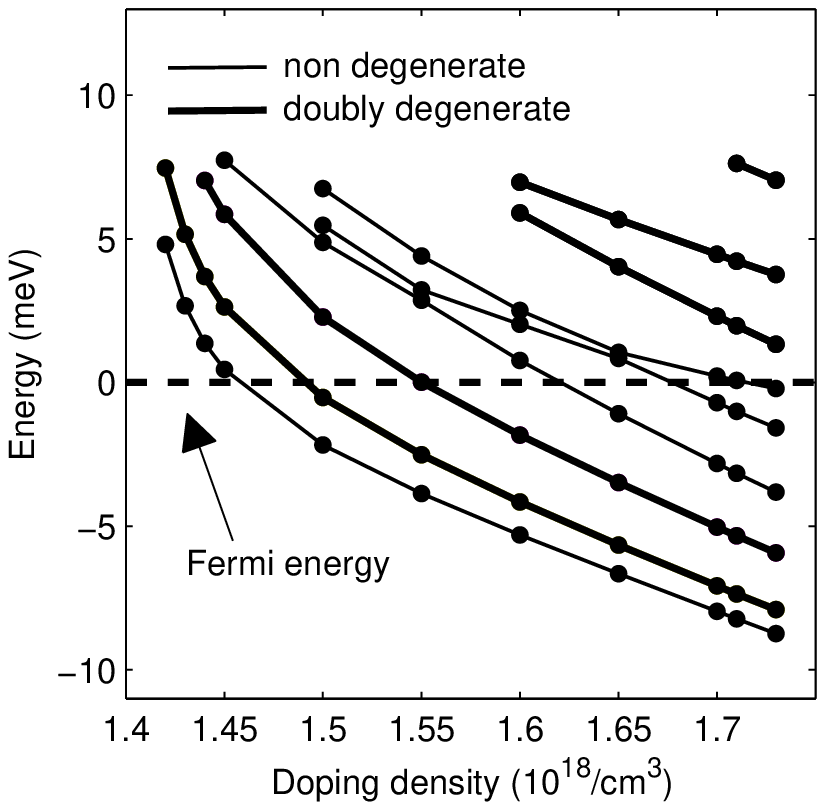}
 \caption{Electron subband energy vs doping density $\rho_D$. Each bullet represents the subband edge resulting from the self-consistent calculation. The lines (thinner for non-degenerate subbands, thicker for doubly degenerate subbands) are a guide to the eyes. }
 \label{electron_subbands}
 \end{figure}

The charge density is obtained from the occupation of an increasing number of energy subbands. The evolution of subband energies with the doping density is shown in Fig.~\ref{electron_subbands} with respect to the Fermi energy. As the doping increases, subbands fall more and more below the Fermi level. Due to the six-fold symmetry, these states are singly or doubly degenerate, as indicated.

\subsection{Valence band holes\label{Holes}}

Nanowires are easily $p-$doped, for example with Si, which is amphoteric for GaAs.\cite{Dufouleur10} In this section we discuss the formation of the hole gas in the same structure as in the previous section. We use a parabolic description with the appropriated effective masses, as discussed in Sec.~\ref{Method}, reported in Tab.~\ref{table_mat_par}. Due to this parabolic model,
the difference between electron and hole gases to be discussed below reside only in the much heavier effective mass of holes.

\begin{figure}
 \includegraphics[width=0.9\textwidth]{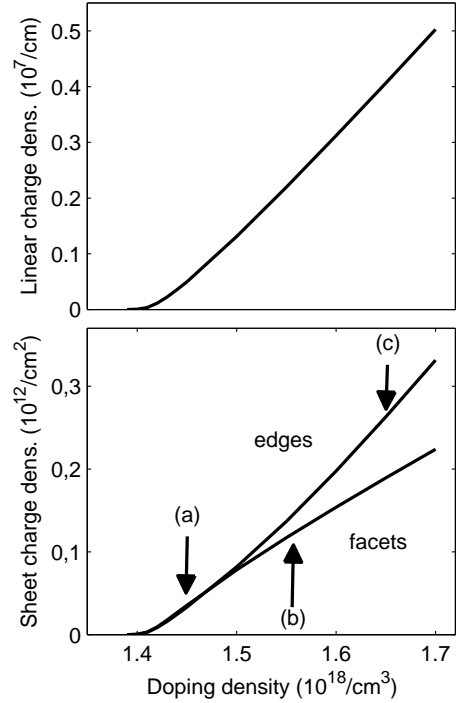}
 \caption{Top: Linear charge density of the free-hole gas as a function of the acceptor density $\rho_A$. Bottom: sheet density of the free-hole gas calculated along the facet-to-facet and edge-to-edge directions indicated in Fig.~\ref{Geometry}. Arrows indicate three different densities shown in in Fig.~\ref{hole_densities}
}
 \label{hole_free_charge_density}
 \end{figure}

Figure \ref{hole_free_charge_density} shows the free charge density of holes as a function of the doping level. Similarly to the electron gas, there is a threshold doping density $~1.4 \times 10^{18} cm^{-3}$, below which no free charge is obtained. Above this threshold the linear charge density (top panel) increases linearly, as for conduction electrons. The equivalent sheet charge density (bottom panel) along the edges-to-edge and facet-to-facet directions [Fig.~\ref{hole_free_charge_density}] of the structure also behave similarly to the case of electrons. In particular, at low doping the two densities are similar, with the sheet charge density at the edges taking over when the doping increases.

\begin{figure}
 \includegraphics[width=0.45\textwidth]{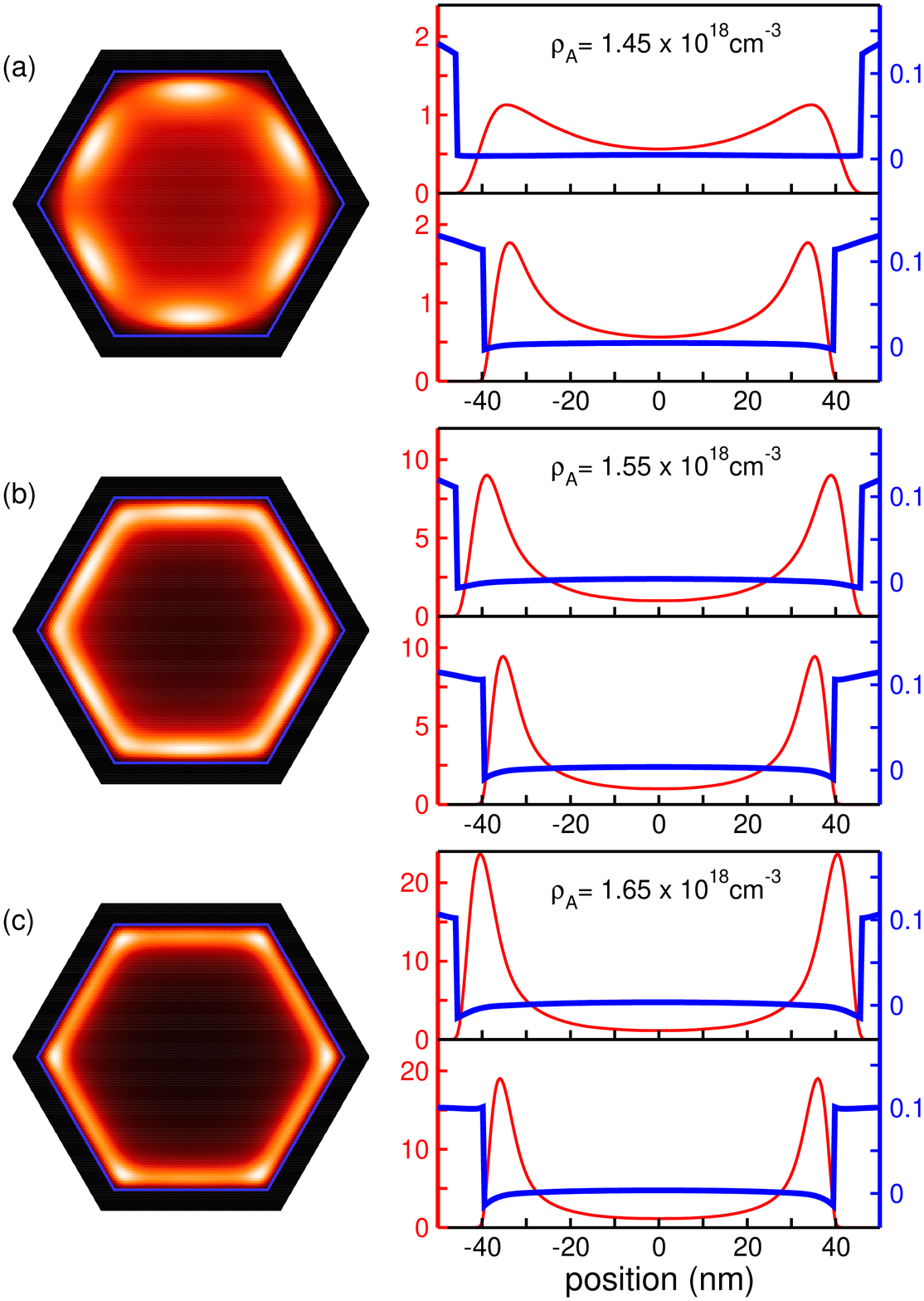}
 \caption{Spatial distribution of the free-hole gas for the three values of the doping density indicated as (a), (b), and (c) in Fig.~\ref{hole_free_charge_density} (bottom panel). For each density we show (left) the 2D map of the charge, with the GaAs/AlGaAs interface indicated as a solid line, and (right) the charge density profile (thin line, left axis, units of $10^{15}$~cm$^{-3}$) and the self-consistent valence-band profile (thick line, right axis, units of eV, $\mu=0$) along the edge-to-edge direction (right, top) and along the facet-to-facet (right, bottom). Note the at the lowest doping (a) the free holes are localized mainly along the facets, contrary to the electron case, and at intermediate doping (b) the free holes form an hexagonal ring, a regime which have no counterpart for conduction electrons. }
 \label{hole_densities}
 \end{figure}

However, it turns out that electrons and holes have a very different behavior and localization with respect to conduction electrons, particularly at low and intermediate charge densities. Figure \ref{hole_densities} shows the free hole density profiles at the three densities (a), (b), and (c) indicated in Fig.~\ref{hole_free_charge_density} (bottom panel). Although the sheet charge density is nearly equal in the two directions at the lowest density, the hole charge density is peaked at the facets, while the distribution along the edge-to-edge direction is low at the edges and broader toward the core. This is sharp contrast with conduction electrons, where at low doping the charge forms a cylindrical distribution in the core.

At intermediate densities [Fig.~\ref{hole_densities}(b)] the gap between the facets is filled, and the hole gas is uniformly distributed at the GaAs/AlGaAs interfaces. Only weak modulations are present along the interface going around the core in this very interesting intermediate regime. Therefore, in this case the hole gas indeed resembles a 2D hole gas which is six-fold bent around the wire. This regime has no counterpart for conduction electrons. At the largest density [Fig.~\ref{hole_densities}(c)], instead, the charge is concentrated at the edges, similarly to the case of electrons.

\begin{figure}
 \includegraphics[width=0.45\textwidth]{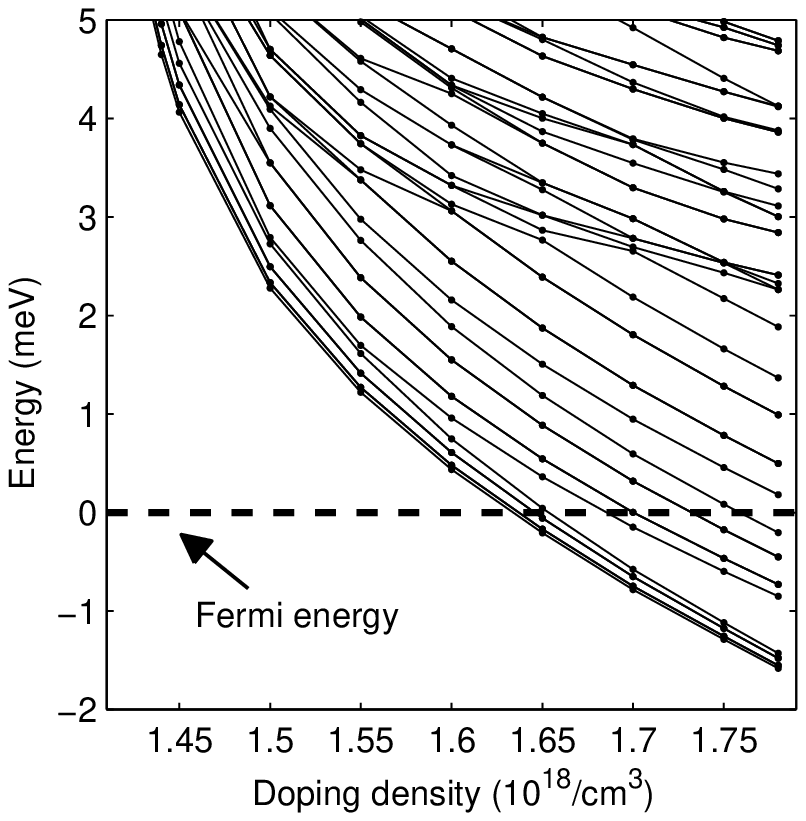}
 \caption{Hole subband energy against doping density $\rho_A$. Each bullet represents the subband threshold resulting from a self-consistent calculation. The lines are a guide to the eyes.
}
 \label{hole_subbands}
 \end{figure}

The evolution of the hole subbands with doping density is shown in Fig.~\ref{hole_subbands}. The behavior is clearly similar to that of conduction electrons, except for the large density of levels due to the larger mass. Two sets of levels can be clearly recognized, corresponding to levels with one additional nodal surface in the radial direction.

\subsection{Origin of localization patterns\label{Patterns}}

To rationalize the behavior of electrons and holes vs doping, particularly at low density, we can consider the following energy contributions: 1) repulsive Coulomb energy between like particles, which requires maximization of the average inter-particle distance and therefore favors localization at the edges of the inner GaAs/AlGaAs interface, 2) \emph{vertical} localization energy, that is the kinetic energy required to localize charge at the GaAs/AlGaAs interfaces, analogously to a planar heterojunction, and 3) \emph{lateral} localization energy, that is the energy required to localize parallel to the interface, along a facet or at an edge, the latter being clearly larger.

At low density, when the Coulomb contribution is small, conduction electrons distribute over the core, since vertical localization energy is large. Lateral localization would be also dominant over the small Coulomb energy in this regime, hence the isotropic, cylindrical symmetry. The charge gets localized near the GaAs/AlGaAs interfaces only if the density and, hence, the average Coulomb energy is so large that localization at the edges is energetically favorable. In this regime, indeed, the small additional cost in lateral localization energy at the edges with respect to facets is overcompensated by gain in Coulomb energy due to inter-edge repulsion. For holes, instead, the small kinetic energy (one order of magnitude smaller than for conduction electrons) makes it easier to localize at the accumulation layer near the interface even at very low densities. Since the Coulomb energy is anyway small in this regime, at the interfaces hole density may localize at the facets, rather than at the edges, where the lateral confinement energy would be larger.

We finally note that the localization depends essentially on the amount of free charge, which in the present calculation is tuned by the doping density. Alternatively, one could fix the doping density to a large value, and deplete the electron (hole) gas using an external gate grown around the wire, recovering the different localization regimes shown above using a negative (positive) voltage applied to it. We have explicitly calculated the charge evolution by simulating an all around gate, and indeed recovered very similar results (not shown). In the next chapter, we will show instead what happens by application of a back-gate to a NW.

\subsection{Biased system \label{Bias}}

\begin{figure}
 \includegraphics[width=0.45\textwidth]{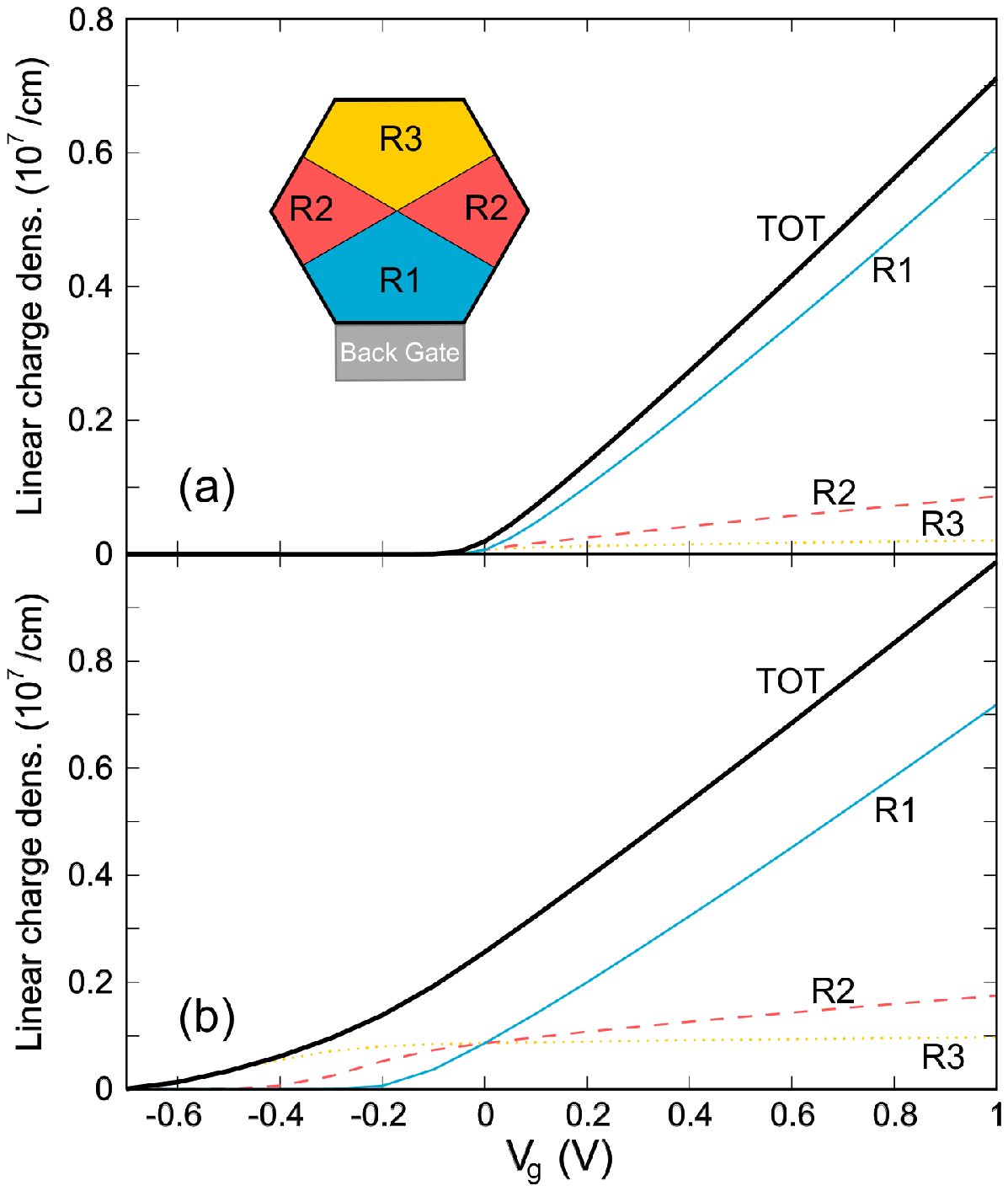}
 \caption{Linear electron density integrated over the three domains of the NW illustrated in the inset, as a function of the gate voltage. Solid blue lines are used for the domain labeled as R1, dashed red lines for R2, and dotted yellow lines for R3. The total electron density is also plotted in thick solid black lines.
Two densities of n-dopants are represented (a) $1.4\times 10^{18} \, \mathrm{cm^{-3}}$ and (b) $1.64\times 10^{18} \, \mathrm{cm^{-3}}$.}
 \label{electron_sector_dens}
 \end{figure}

Transport properties of NWs can be measured by integrating them in planar field effect transistors (FET) \cite{mescherIEEE77,tangACS5}. In such devices a  NW is placed on a substrate formed from a dielectric layer which separates the wire from another layer acting as a global back gate. The applied gate voltage depletes or populates the NWs
depending on the sign, and hence, it can be used to modulate the conductance between the source and drain electrodes. We next simulate the effect of such a back gate assuming that it is in remote contact with one of the external facets of the NW, hereafter referred to as the bottom facet (see inset in Fig.~\ref{electron_sector_dens}). The potential is set to $V=-e V_g$ on the basis edge, where $V_g$ is the applied electrode voltage in the range $V_g= \pm 1$~V.  Since the boundary of our simulation domain corresponds to the CSNW surface and it is at least 60~nm distant from the free carriers, we assume that the possible modulations of $V$ along the edges have little influence on the carrier localization in the core.

\begin{figure}
\includegraphics[width=0.45\textwidth]{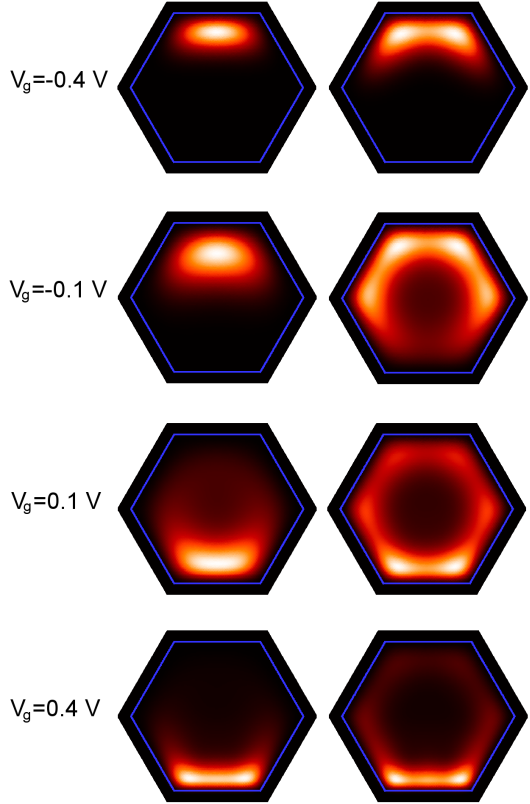}
\caption{Spatial distribution of the free-electron gas in two CSNWs with doping
densities $1.4\times 10^{18} \, \mathrm{cm^{-3}}$ (left) and $1.64\times 10^{18} \,
\mathrm{cm^{-3}}$ (right) at four selected values of the gate voltage.
Depending on the sign of the voltage, the free electrons tend to localize
near the top/bottom parts of the wire.
}
\label{electron_contours}
\end{figure}

In Fig.~\ref{electron_contours} we show the electron charge density profiles at four selected gate
voltages for two NWs with a low (left) and high (right) density of dopants. As anticipated, the charge
density no longer shows the symmetric distributions obtained at zero voltage, and tends to accumulate
at the top and bottom of the wire section for negative and positive voltages respectively. Interestingly,
for the NW with higher density of dopants we observe how the six quasi-1D channels obtained at zero voltage
can be switched to four and two channels by tuning the voltage to $V_g=-0.1\,V$ and $V_g=\pm 0.4\,V$. Also note the different localization for negative voltages at low or high doping, where charge is distributed in a broad channel or in two narrower channels, respectively.

The quantitative effect of the gate voltage can be observed in Fig.~\ref{electron_sector_dens} where
we represent the linear charge density as obtained by integrating the volume charge density over the three domains depicted in the inset, as a function
of the applied voltage. The total linear charge density is also plotted (black solid line). Clearly, the three curves coincide at zero bias due to symmetry. A positive gate voltage does not deplete the top regions of the wire in favor of a larger electronic
concentration in the bottom of the wire. Instead, the charge density raises quasi-linearly in all regions, although this increase is much pronounced in the bottom regions R1 and R2, while in region R3 the charge remains nearly constant at the zero bias value. On the other hand, the evolution of the
charge density is not linear when the NW becomes depleted by applying negative voltages. This fact can be
associated with the smaller density of subbands to populate at such low-density regimes. Finally, we notice
that, despite the different density of dopants, at high positive voltages both systems reach the same
density regime with the electrons concentrated at the two bottom edges of the wire (see, e.g., bottom
plots in Fig~\ref{electron_contours}).

\begin{figure}
 \includegraphics[width=0.45\textwidth]{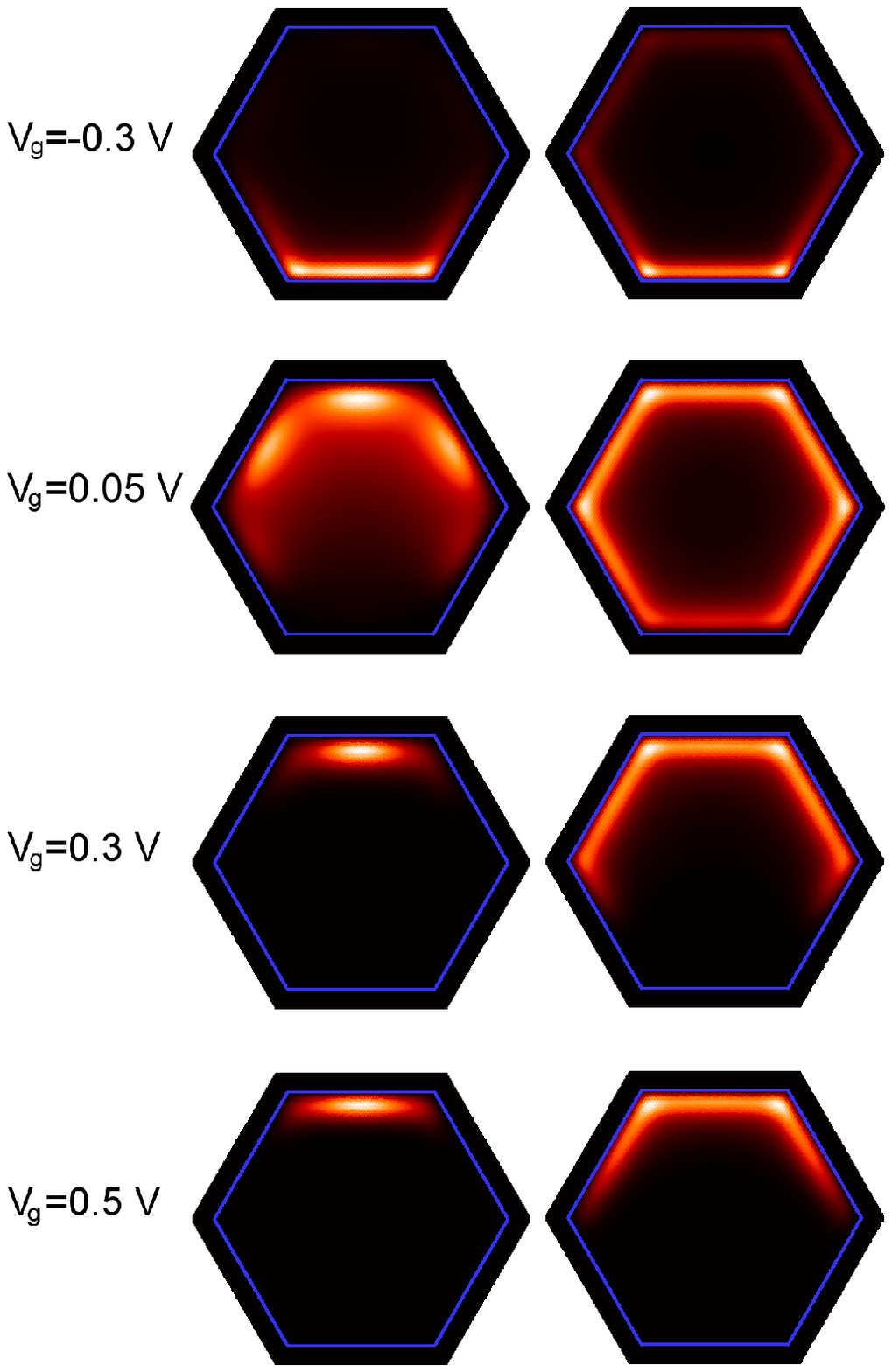}
 \caption{Spatial distribution of the free-hole gas in two CSNWs with doping densities $1.45\times 10^{18} \, \mathrm{cm^{-3}}$ (left) and $1.65\times 10^{18} \, \mathrm{cm^{-3}}$ (right) at four selected values of the gate voltage.
New charge arrangements are observed not present in the electron-gas case, e.g. the charge accumulation at the three top facets of the wire ($V_g=0.05 \, \mathrm{V}$).
}
\label{hole_contours}
 \end{figure}

\begin{figure}
\includegraphics[width=0.45\textwidth]{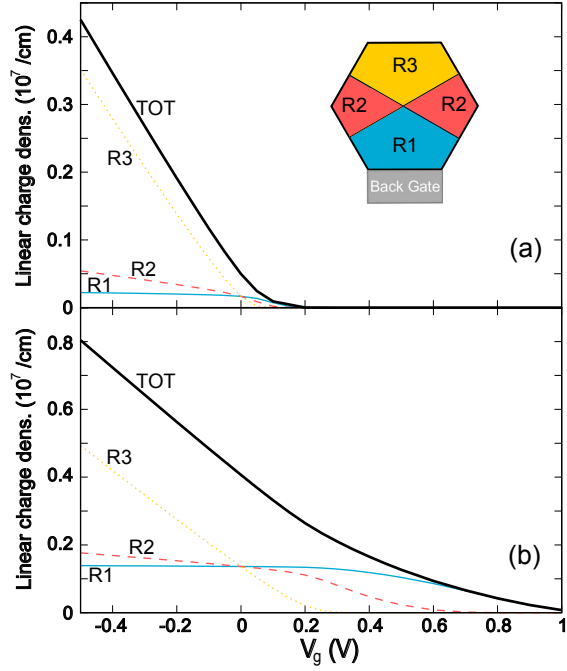}
\caption{Linear hole density integrated over the three domains of the NW illustrated in the inset, as a function of the gate voltage. Solid blue lines are used for the domain labeled as R1, dashed red lines for R2, and dotted yellow lines for R3. The total hole density is also plotted in thick solid black lines. Two densities of p-dopants are represented (a) $1.45\times 10^{18} \, \mathrm{cm^{-3}}$ and (b) $1.65\times 10^{18} \, \mathrm{cm^{-3}}$.}
\label{hole_sector_dens}
\end{figure}

Figures~\ref{hole_contours} and~\ref{hole_sector_dens} show the corresponding results for hole populated NWs with two different densities of dopants. Clearly, the applied voltage induces the opposite effect than in the electronic system, increasing (decreasing) and localizing the charge density at the bottom (top) of the NW at negative (positive) voltages. In parallel with the result obtained at zero voltage, the NW with higher density of $p$-dopants shows qualitatively the same behavior as its $n$-doped counterpart. Thus, in the right panels of Fig.~\ref{hole_contours} it can be seen that, again, a NW with two and four quasi-1D hole channels can be attained by tuning the applied voltage. Nevertheless, the trend of the hole density to localize in the facets at low density regimes enables the observation of new voltage-induced charge arrangements when the density of dopants is lower. For instance, at $V_g=0.05\,V$ one can see that the charge density is concentrated in the center of the three top facets of the NW, while at $V_g=-0.3\, V$ a 2D hole gas is clearly formed over the bottom facet.

To conclude, in Fig.~\ref{hole_sector_dens} we show the effect of the gate voltage on the linear hole density integrated over the three regions of the NW. The results are equivalent to the electronic case, with the role of the three regions reversed due to the opposite sign of the carriers, and can be
rationalized along the same lines.

\section{Conclusions\label{Conclusions}}

We have determined the properties of an electron/hole gas in a realistic model of \emph{radial} heterojunctions formed in a modulation doped GaAs/AlGaAs CSNWs, emphasizing the different localization and symmetry which can be realized as a function of the doping level and/or by a back-gate. Contrary to planar heterojunctions, conduction electrons do not form a 2DEG well localized at the GaAs/AlGaAs interface, but rather show transitions between different charge distributions across the section of the CSNWs. For conduction electrons, these are similar to phases predicted in GaN/AlGaN CSNWs\cite{wongNL11}. For $p$-doped structures, the heavier effective mass result in additional phases for the hole gas which have no counterpart for conduction electrons. In particular, it is possible to form a nearly uniform six-fold bent 2D hole gas at the GaAs/AlGaAs interfaces.

In general, however, our results show that the electron and hole gases in these structures deviate substantially from any isotropic distribution at the GaAs/AlGaAs heterointerfaces. It is particularly important that, depending on the density and possibly the applied gate potential, the charge density reshape between quasi-1D and quasi-2D distributions. This is likely to be reflected in multiple and tunable time-scales of scattering mechanisms.\cite{Masumoto10} In most cases, therefore, we expect complex behavior in possible future transport experiments.

\begin{acknowledgments}
We acknowledge useful discussions with S. Roddaro, H. Strickman, A. Kretinin, I. Zardo, D. Spirkoska, R. Calarco.
We acknowledge partial financial support from MIUR-FIRB RBIN06JB4C and MIUR-PRIN 2008H9ZAZR.
\end{acknowledgments}


\end{document}